\documentclass[journal]{IEEEtran}
\IEEEoverridecommandlockouts
\usepackage{amssymb,amsmath}
\usepackage{amsfonts}
\usepackage{hyperref}
\usepackage[backend=biber]{biblatex}
\usepackage{microtype}
\usepackage{graphics}
\usepackage{color}
\usepackage{tikz}
\usepackage{url}
\usepackage{enumitem}
\DeclareGraphicsExtensions{.pdf,.png,.jpg}

\makeatletter
\allowdisplaybreaks
\makeatother
\usepackage{algorithmic}
\usepackage{array}
\usepackage{textcomp}
\usepackage{stfloats}
\usepackage{verbatim}
\usepackage{graphicx}
\usepackage{balance}
\begin{document}
\title{Transmission Investment Coordination using MILP Lagrange Dual Decomposition and Auxiliary Problem Principle}

\author{Sambuddha~Chakrabarti,~\IEEEmembership{Member,~IEEE,}
        Hosna~Khajeh,~\IEEEmembership{Graduate Student Member,~IEEE,}
        Thomas~R~Nudell,~\IEEEmembership{Member,~IEEE,}
        Mohammad~Reza~Hesamzadeh,~\IEEEmembership{Senior Member,~IEEE,}
        and~Ross~Baldick,~\IEEEmembership{Fellow,~IEEE,}
\thanks{S. Chakrabarti is currently a staff researcher at Princeton University, Princeton, NJ, USA. He was previously a postdoctoral researcher at KTH Royal Institute of Technology, Stockholm, Sweden. H. Khajeh is a doctoral researcher at the School of Technology and Innovations, University of Vaasa, Vaasa, Finland. T. Nudell is the cofounder and CEO of Piq Energy Corp, Union City, CA, USA. M. R. Hesamzadeh is an associate professor at KTH Royal Institute of Technology. R. Baldick is a professor emeritus with the Department of Electrical \& Computer Engineering, The University of Texas at Austin, Austin, TX, USA. This work was supported in parts by Svenska Kraftn{\"a}t \& by National Science Foundation (grant number:ECCS-1406894)}
}


\maketitle
\begin{abstract}
This paper considers the investment coordination problem for the long term transmission capacity expansion in a situation where there are multiple regional Transmission Planners (TPs), each acting in order to maximize the utility in only its own region. In such a setting, any particular TP does not normally have any incentive to cooperate with the neighboring TP(s), although the optimal investment decision of each TP is contingent upon those of the neighboring TPs. A game-theoretic interaction among the TPs does not necessarily lead to this overall social optimum. We, therefore, introduce a social planner and call it the Transmission Planning Coordinator (TPC) whose goal is to attain the optimal possible social welfare for the bigger geographical region. In order to achieve this goal, this paper introduces a new incentive mechanism, based on distributed optimization theory. This incentive mechanism can be viewed as a set of rules of the transmission expansion investment coordination game, set by the social planner TPC, such that, even if the individual TPs act selfishly, it will still lead to the TPC's goal of attaining overall social optimum. Finally, the effectiveness of our approach is demonstrated through several simulation studies.\\
\end{abstract}
\IEEEpeerreviewmaketitle
\section{Notations and Conventions}\label{notations}
\begin{itemize}
    \item \textit{Sets}
    \begin{itemize}[label={}]
        \item $\mathcal{N}$: Set of nodes.
        \item $\mathcal{N}_z$: Set of nodes in region $z$.
        \item $\mathcal{N}_{z'}$: Set of nodes in region $z'$ (i.e. neighbors of $z$).
        \item $G_z$: Set of generators in region $z$.
        \item $G_{zn}$: Set of generators in $z$ connected to node $n$.
        \item $H_z$: Set of existing transmission lines in region $z$.
        \item $H_{zn}$: Set of existing lines in $z$ connected to node $n$.
        \item $\Tilde{H}$: Set of existing lines shared between multiple regions
        \item $K_z$: Set of candidate lines in region $z$.
        \item $D_z$: Set of loads in region $z$.
        \item $D_{zn}$: Set of loads in region $z$ connected to node $n$.
        \item $Z$: Set of regions or sub-networks.
        \item $S$: Set of scenarios.
        \item $\dagger$ used to denote the transpose of a vector or matrix
    \end{itemize}
    \item \textit{Indices}
    \begin{itemize}[label={}]
        \item $n$: Index for the elements of $\mathcal{N}$.
        \item $g$: Index for the elements of $G$.
        \item $h$: Index for the elements of $H$ and $\Tilde{H}$.
        \item $k$: Index for the elements of $K$.
        \item $d$: Index for the elements of $D$.
        \item $z$: Index for the elements of $Z$.
        \item $s$: Index for the elements of $S$.
    \end{itemize}
    \item \textit{Parameters}
    \begin{itemize}[label={}]
        \item $x_{h}$: Reactance of the transmission line $h$.
        \item $\hat{x}_{k}$: Reactance of the candidate transmission line $k$.
        \item $\overline{L}_{h}$: Capacity of the transmission line $h$.
        \item $\hat{L}_{k}$: Maximum expansion capacity for the candidate line $k$.
        \item $\mathbf{A}_{(n,h)}$: $(n,h)^{th}$ element of the node-to-branch incidence matrix $A$. The element is 1 when node $n$ is the sending end of the line $h$, -1 when it is the receiving end, and 0 otherwise.
        \item $w_s$: Weight of scenario, $s$.
        \item $T_k$: Lifetime of the candidate line $k$.
        \item $r$: Interest rate.
        \item $P_{sd}$: Demand of the $d^{th}$ load in the scenario $s$.
        \item $\underline{P}_{sg}$, $\overline{P}_{sg}$: Lower and upper generating limits of generator $g$ in scenario $s$.
        \item $i_k$, $j_k$: Sending and receiving ends of candidate line $k$ respectively.
        \item $\eta,\gamma,\delta_{APP}$: First and second penalty parameters and the step-length parameter of the APP iterations, respectively.
    \end{itemize}
    \item \textit{Variables \& Functions}
    \begin{itemize}[label={}]
        \item $P_{sg}$: Power generation of generator $g$ in scenario $s$.
        \item $\theta_{sn}$: Node voltage phase angle of node $n$ in scenario $s$.
        \item $u_{zk}$: Binary decision variable with $0$ for no capacity expansion and $1$ for capacity expansion to the maximum limit for the candidate line $k$ in region $z$. 
        \item $\hat{P}_{sk}$: Power flow on candidate line $k$ in scenario $s$.
        \item $P_{sh}$: Power flow on existing line $h$ in scenario $s$.
        \item $C_g(P_{sg})$: Cost function of the generator $g$ (typically convex quadratic or convex cubic), reflecting the fuel cost and heat rate, for producing power, $P_{sg}$.
        \item $C_k(\hat{L}_k)$: Cost of building a new line $k$ of capacity $\hat{L}_k$.
        \item $\lambda^{\sigma}_{zz'i(h)},\lambda^{\sigma}_{zz'i(h)}$: Dual variable in the Auxiliary Problem Principle (APP) iterations for consensus on flow value on the line $h$ shared between regions $z$ and $z'$ at the iteration count $\sigma$.
    \end{itemize}
\end{itemize}

\section{Introduction}\label{introduction}
\IEEEPARstart{T}{he} contemporary electric power network is a complex, large, and dynamically engineered system. It is old, with the average age of transmission assets nearly forty years old. Yet it is constantly evolving to accommodate growing demand and also to incorporate the new technologies like renewable generation, modern communication, and demand-response equipment. The consideration of competitive market setting for the expansion planning problems makes it even more complex \cite{Sambuddha2010}. As such, the expansions of both the generation, as well as the transmission infrastructures are of paramount importance in the long-term planning of power systems across the world. In this paper, we focus on the problem of transmission investment, particularly in the situation where there are multiple transmission planners who do not necessarily have any incentive to cooperate with each other. 
\subsection{Literature Review}
\noindent Previous work has addressed the generation and transmission expansion planning problem in a centralized manner. These studies mainly concentrated on how to develop generation capacities over a specific time horizon. For example, \cite{eia2014electricity} assessed how the electric power industry can change and develop its mix of generating capacities over time according to the future demand, electricity and fuel prices, future regulations, and technological costs. In another similar work introduced by \cite{epa2018documentation}, U.S. Environmental Protection Agency (EPA) tried to provide optimal options for the generation capacity expansion problem. The options include utilizing the demand-side participation, renewable resources, and traditional generation capacities in the future. Reference \cite{young2018us} also considered the effects of customer choices and end-user services on the energy demand and therefore on the generation capacity expansion model. In another work introduced by \cite{ho2021regional} and \cite{mai2013resource}, a capacity expansion model is solved through three different modules. The first module is named supply module, which aims to obtain the cost-efficient investment and operation of electrical sectors. The second module is a demand module trying to maximize the levels of investment on end-user devices, while the third module analyzes the optimal values and key parameters of renewable generators. A few modules written in Python such as “Switch” also provide a thorough generation expansion planning method with regard to the existing dispatch and investment rules for different technologies. These technologies can include storage and demand response resources \cite{johnston2019switch}. 

In a few studies, the authors focused on the problem of transmission and generation expansion planning, simultaneously. For instance, PyPSA has worked on transmission expansion based on integer decision variables and big-M relaxations \cite{brown2017pypsa}. Moreover, GenX model defines a generation expansion planning model considering both investment on centralized and distributed generation capacities, storage-based and demand-side technologies \cite{jenkins2017enhanced} and \cite{jenkins2018electricity}. Additionally, the model takes into account the transmission network expansion decisions as well, but on a regional aggregated or commercially significant constraint basis. In all these studies, the problem of transmission and generation expansion planning has been defined in a centralized fashion and prescriptive mode. 

In the context of real-world transmission networks, the problem of generation and transmission expansion is seldom solved in a centralized manner since the fundamental nature of the problem is that of decentralized decision-making. In reality, there are multiple regions and each region has its separate transmission and generation planner. Each transmission planner is in charge of solving an expansion planning problem in its own region. However, the previous literature did not address the issue of ``how to realize the prescriptive expansion recommendations or what kind of incentivizing schemes to design in order to convert their results to reality." There have been problems of similar nature and mathematical structure to that of our present problem, which has been solved for different
application domains. For instance, in the area of the Internet of Things (IoT),
IoT devices cooperatively execute tasks in a decentralized way
\cite{hou2021incentive}, or \cite{zeng2021novel} put forward an incentive mechanism in which
vehicle users deploy a non-cooperative game to create service
requests, and the system accordingly guides the best available
offloading strategy. Reference \cite{wang2017survey} suggested the decentralized routing
way for Internet Energy which maintains a certain utilization
rate and facilitates fast learning from the mechanism.

To resolve the multi-regional issue, some research discussed the planning problem with multiple transmission planners (TPs) in different regions using game-based approaches. However, the main issue related to these approaches is that the problem may not lead to optimal points for all of the TPs. In this context, authors of \cite{tohidi2014multi} have solved the multi-region transmission expansion planning problem by applying the non-cooperative game-theoretic model in a bi-level optimization formulation, through the use of Nash Equilibrium (NE) solution concept. As illustrated in that paper, the outcome of such an approach typically happens to be sub-optimal compared to the centralized solution where it is assumed that all the different transmission planners (TPs) merge into one entity. The authors also alluded to different compensation mechanisms in that reference which can improve the results. In \cite{tohidi2013free}, the authors have mentioned the interesting effect of ``free-riding," in the context of such non-cooperative, multi-regional transmission planning, which is the presence of entities that make benefit out of expansion by others, without themselves contributing anything. The authors have studied the impact of this free-riding on congestion revenues. In the reference \cite{tohidi2013multi}, the authors have tried to improve the social welfare as presented in \cite{tohidi2014multi} by adopting different solution approaches, whereas in \cite{tohidi2014multiwind}, the multi-regional transmission planing has been explored in the presence of wind generation and its associated uncertainty. References \cite{tohidi2017coordination}, \cite{tohidi2015reactive}, and \cite{tohidi2017sequential} have explored the transmission expansion planning in the presence of generation expansion planning considering pro-active and reactive coordination methods. Reference \cite{tohidi2016optimal} is a good comprehensive collection of the work that has been performed so far in this direction.

It is important to observe that consideration of contingencies and pre- and post-contingency line flows are important metrics for the solution of the transmission capacity expansion problem. For obvious reasons, the inclusion of contingencies increases the computational complexity of the problem. References  \cite{majidiStoch}, \cite{majidiDecomposition}, \cite{majidiIntegration}, and \cite{majidiReducing} explored the issue of inclusion of contingencies in the context of both the deterministic and stochastic settings as well as prescribed ways to eliminate umbrella contingency scenarios and expedite the solution through decomposition methods in their papers. \\
Closely aligned to this problem is the problem of distributed optimization in networked cyber-physical system, in an environment of very limited information and bandwidth, which is presented in the work, \cite{magnusson2017bandwidth}. Although the previous literature mostly results in sub-optimal coordination outcome, this paper introduces a Transmission Planning Coordinator (TPC) that guides Transmission Planners (TP) with a view towards a near global social optimum, as much as possible . It is to be noted, however though, that achieving global optimality is not the primal aim of this work (this is so because, this problem has integer decision variables and is solved in a decentralized manner. Therefore, in principle, achieving global optimality is presumably impossible); however, this paper tries to:
\begin{itemize}
    \item Simulate how transmission expansion works in real-world multi-region settings.
    \item Come up with a market mechanism that presumably works better in comparison to non-cooperative game-theoretic interactions among the regions (in the absence of a social planner)
\end{itemize}
Hence, in the present case, we will assume the existence of a central rule-making agent or market overseeing authority, who sets the rules of the game, and the individual transmission and generation asset owners, who independently solve their own optimization sub-problem.This idea has also been discussed recently by the Federal Energy Regulatory Commission (FERC) at their staff-led workshop \footnote{See:\url{https://www.ferc.gov/news-events/events/staff-led-workshop-establishing-interregional-transfer-capability-transmission}}.
We apply the ideas from the world of distributed optimization and message-passing frameworks to motivate the setting up of such a mechanism design scheme.
Introducing the TPC also unlocks a number of new mechanisms to improve social outcomes, including enabling truth-telling behavior among participating regions based on Vickrey-Clarke-Groves auction and the generalization to Vickrey-Clarke-Groves mechanism. This paper does not elaborate on Vikrey-Clarke-Groves auction and/or Vickrey-Clarke-Groves mechanism but takes the necessary step of introducing the TPC to enable further research. In our work, we have adopted the simplest 50-50 cost sharing of the new shared transmission lines to be constructed between two regions. Adopting the VCG mechanism for inducing truth-telling behavior implies modification of this cost-sharing and cost allocation to each region in a way to capture the perceived utilities of all the other regions, by each region. This is proven to induce truth-telling behavior among participating regions when it comes to revealing their real preference as to whether or not to build the shared transmission lines. Our present work opens up avenues for such future work.

In order to motivate the search for an algorithmic framework, based on which we design our incentive mechanism, we observe the following characteristics of the present optimization problem:\\
\begin{itemize}
    \item The centralized optimization problem is a mixed integer programming problem (MIP) since it has discrete binary decision variables as well as continuous ones. The binary variables decide whether to build a particular line or not along a candidate line, and continuous variables pertaining the generator dispatch and voltage phase angles.
    \item The bigger problem is split into several regions and the candidate pathways.  Moreover, some existing lines are shared between two different regions.
    \item The decision variables corresponding to the shared lines are also shared among each region and there needs to be consensus among the values decided by the different regions.
\end{itemize}
In order to satisfy the above requirements, we propose a two staged incentive mechanism design, which consists of the following stages or layers:
\begin{itemize}
    \item The first stage, based on Distributed Stochastic Optimization, solves for the integer variables and approximately solves for the continuous variables.
    \item The second stage, based on Auxiliary Problem Principle (APP), solves for the continuous variables more accurately, once the discrete variable values are determined.
\end{itemize}
For the first stage, we use the algorithm introduced in \cite{aravena2017distributed} for solving Stochastic Unit Commitment problem in a distributed manner. It is an asynchronous distributed algorithm for maximizing the Langrangian dual problem of the classic Unit Commitment (UC) formulation. The synchronous version of the algorithm was presented in the earlier works \cite{papavasiliou2011reserve}, \cite{papavasiliou2015applying}, and \cite{papavasiliou2013multiarea}. In those papers, the authors have determined the commitment decisions, which are binary discrete variables, of slow responding generators across different scenarios, through the use of ``non-anticipativity constraints." Instead of a scenario decomposition, in our work, we will be performing a region decomposition.
For the second stage of the mechanism design, we will be making use of the Auxiliary Problem Principle (APP), which was introduced by Cohen \emph{et al.} in the seminal works \cite{cohen1978optimization} and \cite{cohen1980auxiliary}. It was subsequently used in power flow problems for coarse decomposition, both for a region decomposition in \cite{baldick1999fast}, \cite{kim1997coarse}, \cite{kim2000comparison}, and \cite{ebrahimian2000state} and for decomposition across different dispatch intervals and contingency scenarios in \cite{Sambuddha2017}, \cite{chakrabarti2019look}, \cite{chakrabarti2020look}.\\
Reflecting on what was stated, the novelty of our paper can be summarized as follows:
\begin{itemize}
    \item Previous approaches have taken a non-cooperative game theoretic approach. This results in sub-optimal coordination outcomes. By
    introducing a TPC, individual TPs are incentivized to converge towards the near global social optimum while still following their locally selfish objectives. This
    approach has not, to our knowledge, been proposed in the context of
    inter-regional transmission planning.
    \item Although this paper has been inspired by the BCD method, it deals with a totally different problem (transmission expansion planning problem) and the algorithm has been completely modified to be compliant with the new application.
    \item We propose the novel application of APP in which TPs reach a consensus regarding the flow and voltage angles of inter-regional shared lines between them.
\end{itemize}
The rest of the current paper is organized as follows:  In Section \ref{centralized} we present the centralized transmission expansion planning for reference and for the sake of completeness. Section \ref{Distributed} presents the mathematical formulation of both the stages of our distributed algorithmic incentive mechanism design and the steps thereof. In Subsection \ref{stage_I_mech_model}, we will introduce the first stage of the mechanism with its mathematical model. In Subsection \ref{mech_steps}, we mention the steps of the first stage incentive mechanism design. Subsection \ref{APP_mech} details the second stage mechanism design model, which is based on the APP algorithm, while Subsection \ref{second_steps} walks us through the steps of the second stage mechanism design. We state the results of numerical simulations in Section \ref{numerical} and in Section \ref{conclusion}, we draw some concluding remarks, while pointing to the future research directions.

\section{Centralized Coordination (Benchmark)}\label{centralized}
The centralized coordination of transmission investment is modeled in (\ref{Step2:General_one}) to create a benchmark for the distributed coordination algorithm. In the centralized version, the TPC minimizes the total operations and investment costs in all regions. 
The model presented here is a slight modification of the one presented in \cite{tohidi2014multi}:\\
\begin{subequations}\label{Step2:General_one}
\begin{gather}
\min_{P_{sg}, u_{zk}}\sum_{z\in Z}\sum_{s\in S}\sum_{g\in{G}_z}w_sC_{g}(P_{sg})\notag\\+\sum_{z\in Z}\sum_{k\in K_z}u_{zk}C_{k}({\hat{L}}_{k})\frac{r(1+r)^{T_k}}{(1+r)^{T_k}-1}\label{first_Step2:General_one}\\
\mbox{Subject\;to:\:}\sum_{g\in G_{zn}}P_{sg}-\sum_{d\in D_{zn}}P_{sd}=\notag\\\sum_{h\in H_{zn}}\mathbf{A}_{(n,h)}P_{sh}+\sum_{k\in{K_z},i_k=n}\hat{P}_{sk}-\sum_{k\in{K_z},j_k=n}\hat{P}_{sk} \notag\\\forall{z\in Z},\forall{s\in S},\forall{{n}\in\mathcal{N}_z}\label{second_Step2:General_one}\\
P_{sh}=\frac{1}{x_h}\sum_{n\in \mathcal{N}}\mathbf{A}_{(n,h)}\theta_{sn},\forall{z\in Z},\forall{s\in S},\forall{{h}\in H_z}\label{secondI_Step2:General}\\
\hat{P}_{sk}=u_{zk}\frac{1}{\hat{x}_k}(\theta_{si}-\theta_{sj}),k \rightarrow (i,j)\notag\\\forall{z\in Z},\forall{s\in S},\forall{{k}\in K_z}\label{secondII_Step2:General}\\
-{\overline{L}}_{h}\leq P_{sh}\leq{{\overline{L}}_{h}},\;\forall{z\in Z},\forall{s\in S},\forall{{h}\in H_z}\notag\\\label{third_Step2:General_one}\\
-{u_{zk}{\hat{L}}_{k}}\leq \hat{P}_{sk}\leq{u_{zk}{\hat{L}}_{k}},\;\forall{z\in Z},\forall{s\in S},\forall{{k}\in K_z}\notag\\\label{third_Step2II:General}\\
{\underline{P}}_{sg}\leq P_{sg}\leq{{\overline{P}}_{sg}},\;\forall{z\in Z},\forall{s\in S},\forall{g\in{G_z}}\label{fourth_Step2:General}\\
u_{zk}\in\{0,1\},\;\forall{z\in Z},\forall{{k}\in K_z},\;\label{fifth_Step2:General}
\end{gather}
\end{subequations}
In the objective function, the first term is the expected operational cost and the second term is the total investment amount taking into account the interest of the transmission investment cost. 
Constraint (\ref{second_Step2:General_one}) represents the power balance for each node. 
Constraints (\ref{secondI_Step2:General}) and (\ref{secondII_Step2:General}) define line flows for the existing lines and the candidate lines, respectively, while (\ref{third_Step2:General_one}) and (\ref{third_Step2II:General}) define the line flow limits for existing and candidate lines, respectively. 
The last two constraints are for generation limits and for permissible values that the binary variables for constructing new lines can assume.

\section{Distributed Coordination Mechanism}\label{Distributed}
In this section, we present the design of a novel two-stage incentive mechanism. The first stage is based on, and inspired by the Distributed Stochastic Mixed Integer Optimization and the concept of scenario decomposition. 
We extend the idea of a ``scenario decomposition" to ``area decomposition." 
The outcome of the first stage determines the investment decisions as to whether or not and where to invest in building new transmission lines. It also gives us the approximate values of tie-line flows and generation values.
\subsection{Mathematical Model of Stage I Mechanism Design}\label{stage_I_mech_model}
For the first stage, we present an algorithmic incentive mechanism design based on the asynchronous distributed algorithm presented by the authors in \cite{aravena2017distributed}. 
In constraint (\ref{fourth_Step21:General}), $M$ is a large number. The purpose of this constraint is to linearize the constraint (\ref{secondII_Step2:General}) in Section \ref{centralized}, which is non-linear due to the presence of product terms.
\begin{subequations}\label{Step2:General_one_Area}
\begin{gather}
\min_{P^a_{g_q}}\sum_{z\in Z}\Bigg(\sum_{s\in S}\sum_{g\in{G_z}}w_sC_{g}(P_{sg})\notag\\+\sum_{k\in K_z}\Big\{u_{zk}C_k({\hat{L}}_k)\frac{r(1+r)^{T_k}}{(1+r)^{T_k}-1}\Big\}\Bigg)\label{first_Step2:General}\\
\mbox{Subject\;to:\:}\sum_{g\in G_{zn}}P_{sg}-\sum_{d\in D_{zn}}P_{sd}=\notag\\\sum_{h\in H_{zn}}\mathbf{A}_{(n,h)}P_{sh}+\sum_{k\in{K_z},i_k=n}\hat{P}_{sk}-\sum_{k\in{K_z},j_k=n}\hat{P}_{sk} \notag\\\forall{z\in Z},\forall{s\in S},\forall{{n}\in\mathcal{N}_z}\label{second_Step2:General}\\
P_{sh}=\frac{1}{x_h}\sum_{n\in \mathcal{N}}\mathbf{A}_{(n,h)}\theta_{sn},\forall{z\in Z},\forall{s\in S},\forall{{h}\in H_z}\label{third_Step2:General}\\
-M(1-u_{zk})\leq\hat{P}_{sk}-\frac{1}{\hat{x}_k}(\theta_{si}-\theta_{sj})\leq M(1-u_{zk})\notag\\k \leftarrow (i,j),\forall{z\in Z},\forall{s\in S},\forall{{k}\in K_z}\label{fourth_Step21:General}\\
-{\overline{L}}_{h}\leq P_{sh}\leq{{\overline{L}}_{h}},\;\forall{z\in Z},\forall{s\in S},\forall{{h}\in H_z}\notag\\\label{fifth_Step2:General}\\
-{u_{zk}{\hat{L}}_{k}}\leq \hat{P}_{sk}\leq{u_{zk}{\hat{L}}_{k}},\;\forall{z\in Z},\forall{s\in S},\forall{{k}\in K_z}\notag\\\label{sixth_Step2:General}\\
{\underline{P}}_{sg}\leq P_{sg}\leq{{\overline{P}}_{sg}},\;\forall{z\in Z},\forall{s\in S},\forall{g\in{G_z}}\label{sixth_Step2I:General}\\
u_{zk}\in\{0,1\},\;\forall{z\in Z},\forall{{k}\in K_z},\;\label{sixth_Step2II:General}\\
\text{Non-Anticipativity Constraints}\notag\\
u_{zk}=u_{k}\leftrightarrow \pi_{zk},\;\forall k\in K\label{seventh_Step2:General}\\
|\hat{P}_{zsk}|=\hat{\mathbb{P}}_{sk}\leftrightarrow \mu_{zsk},\;\forall k\in K\label{seventh_Step3:General}\\
|{P}_{zsh}|=\mathbb{P}_{sh}\leftrightarrow \mu_{zsh},\;\forall h\in \Tilde{H}\label{seventh_Step4:General}\\
\theta_{zsi(k)}=\phi_{si(k)}\leftrightarrow \xi_{zsi(k)},\;\forall k\in K\label{seventh_Step5:General}\\
\theta_{zsj(k)}=\phi_{sj(k)}\leftrightarrow \xi_{zsj(k)},\;\forall k\in K\label{seventh_Step6:General}\\
\theta_{zsi(h)}=\phi_{si(h)}\leftrightarrow \xi_{zsi(h)},\;\forall h\in \Tilde{H}\label{seventh_Step7:General}\\
\theta_{zsj(h)}=\phi_{sj(h)}\leftrightarrow \xi_{zsj(h)},\;\forall h\in \Tilde{H}\label{seventh_Step8:General}\\
\mathbb{P}_{sh}=\frac{1}{x_h}(\phi_{si(h)}-\phi_{sj(h)}),\forall{s\in S},\forall{{h}\in \Tilde{H}}\label{third_Step9:General}\\
-M(1-u_{k})\leq\hat{\mathbb{P}}_{sk}-\frac{1}{\hat{x}_k}(\phi_{si(k)}-\phi_{sj(k)})\leq M(1-u_{k})\notag\\u_k\in\{0,1\},k \leftarrow (i,j),\forall{s\in S},\forall{{k}\in K}\label{third_Step10:General}
\end{gather}
\end{subequations}
The region-decomposed Lagrangian dual of the above optimization problem is stated below in mathematical expressions (\ref{zero_area_decomposition})--(\ref{second_area_decomposition}). 
This set of problems is solved by the different TPs while being coordinated by the TPC. The penalty cost appearing to the right of $\leftrightarrow$ plays the pivotal role of 
motivating the TPs to perform such calculations and report the updated variable values to the TPC.
\begin{subequations}
\begin{gather}
\max_{\mathbf{\pi},\mathbf{\mu},\mathbf{\xi}}f_0(\mathbf{\pi},\mathbf{\mu},\mathbf{\xi})+\sum_{z\in Z}f_z(\mathbf{\pi},\mathbf{\mu},\mathbf{\xi}),\;\text{where}\label{zero_area_decomposition}\\
f_0(\mathbf{\pi},\mathbf{\mu},\mathbf{\xi})\notag\\=\inf_{u_k,\hat{\mathbb{P}}_{sk},\mathbb{P}_{sh},\phi_k,\phi_h}\Bigg(-\sum_{z\in Z}\sum_{k\in K}\{u_k\pi_{zk}+\sum_{s\in S}(\hat{\mathbb{P}}_{sk}\mu_{zsk}\notag\\+\phi_{si(k)}\xi_{zsi(k)}+\phi_{sj(k)}\xi_{zsj(k)})\}-\sum_{z\in Z}\sum_{h\in\Tilde{H}}\sum_{s\in S}\{\mathbb{P}_{sh}\mu_{zsh}\notag\\+\phi_{si(h)}\xi_{zsi(h)}+\phi_{sj(h)} \xi_{zsj(h)}\}:\text{Subject to: }(\ref{third_Step9:General})-(\ref{third_Step10:General})\Bigg)\label{first_area_decomposition}\\
f_z(\mathbf{\pi},\mathbf{\mu},\mathbf{\xi})=\inf_{P_g, u_{zk}}\Bigg(\sum_{s\in S}\sum_{g\in{G_z}}w_sC_{g}(P_{sg})+\notag\\\sum_{k\in K_z}\Big\{u_{zk}C_k({\hat{L}}_k)\frac{r(1+r)^{T_k}}{(1+r)^{T_k}-1}+u_{zk}\pi_{zk}\notag\\+\sum_{s\in S}(|\hat{P}_{zsk}|\mu_{zsk}+\theta_{zsi(k)}\xi_{zsi(k)}+\theta_{zsj(k)}\xi_{zsj(k)})\Big\}\notag\\+\sum_{s\in S}\sum_{h\in \Tilde{H}}\Bigg\{|{P}_{zsh}|\mu_{zsh}+\theta_{zsi(h)}\xi_{zsi(h)}+\theta_{zsj(h)}\xi_{zsj(h)}\Bigg\}\notag\\:\text{Subject to: } (\ref{second_Step2:General})-(\ref{sixth_Step2II:General})\Bigg)\label{second_area_decomposition}
\end{gather}
\end{subequations}
\begin{figure}[h]
\begin{center}
\includegraphics[scale=0.8]{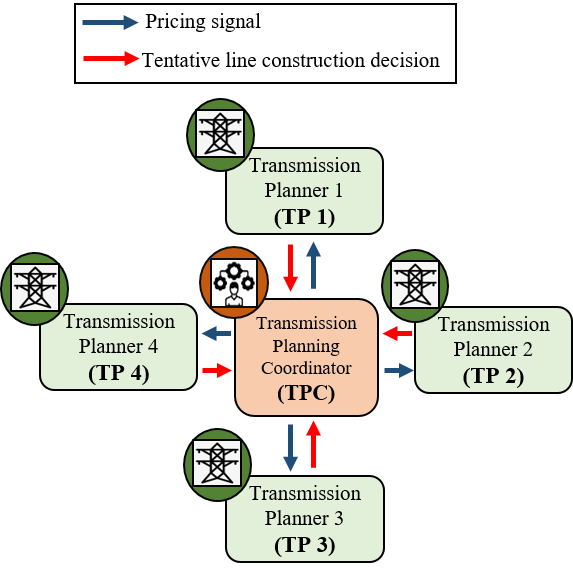}
\caption{Illustration of the incentive mechanism as applied to transmission expansion planning}
\label{fig:SUCTranExpan}
\vspace{-5mm}
\end{center}
\end{figure}
If all the generator cost curves are considered as linear or piece-wise linear, then both (\ref{first_area_decomposition}) and (\ref{second_area_decomposition}) are simply Mixed Integer Linear Programming (MILP) problems, for which there exist standard algorithms for solving. 
\subsection{Steps of Stage I Mechanism Design}\label{mech_steps}
We will now state the steps of our algorithmic incentive mechanism. These are very similar to the ones that appear in \cite{aravena2017distributed}, with the only significant differences being instead of scenario decomposition, we do a region or area decomposition here.
In our mechanism formulation, we will apply the Block Coordinate Descent (BCD) method to reach the consensus regarding the decision to build new transmission lines. The following are the steps of the algorithmic incentive mechanism design:
\begin{itemize}
    \item Each TP of region $z$ evaluates its own $f_z(\pi^{\nu(z)}, \mu_{s}^{\nu(z)}, \xi_{s}^{\nu(z)})$ based on its current updated rewards/penalties for consensus that it receives from the TPC.
    \item Each TP passes to the TPC, the calculated optimizers from its own optimization subproblem. The TPC evaluates $f_0(\tilde{\pi}_z, \tilde{\mu}_{zs}, \tilde{\xi}_{zs})$, $u^{*}_k,\hat{\mathbb{P}}^{*}_{sk},\mathbb{P}^{*}_{sh},\phi^{*}_{si(k)},\phi^{*}_{sj(k)},\phi^{*}_{si(h)},\phi^{*}_{sj(h)}$ (which are the minimizers of  $f_0(\tilde{\pi}_z, \tilde{\mu}_{zs}, \tilde{\xi}_{zs})$), $f_z(\pi^{\nu(z)},\mu_{s}^{\nu(z)},\xi_{s}^{\nu(z)})$, $P^{*}_{sg}$, $u^{*}_{zk}$.
    \item Based on the optimal decision variable values from the last step, the TPC updates the rewards/penalties according to:
    \begin{itemize}[label={}]
        \item $\pi^{\nu(z)+1}=\pi^{\nu(z)}+\frac{\tilde{\alpha}^{\nu}}{\tilde{\beta}_z}(u^{*}_{zk}-u^{*}_k)$.
        \item $\mu_{s}^{\nu(z)+1}=\mu_{s}^{\nu(z)}+\frac{\tilde{\alpha}^{\nu}}{\tilde{\beta}_z}(|\hat{P}^{*}_{zsk}|-\hat{\mathbb{P}}^{*}_{sk})$.
        \item $\mu_{s}^{\nu(z)+1}=\mu_{s}^{\nu(z)}+\frac{\tilde{\alpha}^{\nu}}{\tilde{\beta}_z}(|P^{*}_{zsh}|-\mathbb{P}^{*}_{sh})$.
        \item $\xi_{s}^{\nu(z)+1}=\xi_{s}^{\nu(z)}+\frac{\tilde{\alpha}^{\nu}}{\tilde{\beta}_z}(\theta^{*}_{zsi(k)}-\phi^{*}_{si(k)})$.
        \item $\xi_{s}^{\nu(z)+1}=\xi_{s}^{\nu(z)}+\frac{\tilde{\alpha}^{\nu}}{\tilde{\beta}_z}(\theta^{*}_{zsj(k)}-\phi^{*}_{sj(k)})$.
        \item $\xi_{s}^{\nu(z)+1}=\xi_{s}^{\nu(z)}+\frac{\tilde{\alpha}^{\nu}}{\tilde{\beta}_z}(\theta^{*}_{zsi(h)}-\phi^{*}_{si(h)})$.
        \item $\xi_{s}^{\nu(z)+1}=\xi_{s}^{\nu(z)}+\frac{\tilde{\alpha}^{\nu}}{\tilde{\beta}_z}(\theta^{*}_{zsj(h)}-\phi^{*}_{sj(h)})$.
    \end{itemize}
    \item Each TP computes a lower bound on its objective function by relaxing the integer variables and passes on this information to the TPC, which also computes its lower bound. It then adds up all the regional lower bounds for the current iteration as follows:\\ $LB^{new}=LB_0(\tilde{\pi}_z,\tilde{\mu}_{zs},\tilde{\xi}_{zs})+\sum_{z\in Z}LB_z(\pi^{\nu(z)},\mu_{s}^{\nu(z)},\xi_{s}^{(z)})$.
    \item Each TP updates the above and passes them to the TPC, in response to which, the TPC does the following:\\
    \begin{itemize}[label={}]
        \item \textbf {Step 1:} updates values of rewards/penalties (dual variables) $\pi^{\nu(z)+1}$, $\pi^{\nu(z)}$, $\mu_{s}^{\nu(z)+1}$, $\mu_{s}^{\nu(z)}$, $\xi_{s}^{\nu(z)+1}$, $\xi_{s}^{\nu(z)}$.
        \item \textbf{Step 2:} updates iteration count $\nu(z):=\nu(z)+1$ 
        \item \textbf{Step 3:} calculates new lower bound of objective.
        \item \textbf{Step 4:} calculates $f_{0}(\pi^{\nu(z)},\mu_s^{\nu(z)},\xi_s^{\nu(z)})$
    \end{itemize}
    \item This step is for the primal recovery. In this step, each TP fixes the right hand side of the non-anticipativity constraints to the optimal values of the left-side, like $u_k=u^*_{zk}$ from the last iteration and evaluates the optimization problem for only the TPC and its own region. Let's call the optimum, for each region as upper bound and indicate it as $UB_z$ and for the TPC, as $UB^{TPC}_z$. It then conveys this information to the TPC.
    \item The TPC determines the global $UB$ as $UB=\text{min}\{UB^{TPC}_z\}+\sum_{z\in Z}UB_z$
    \item The TPC terminates the algorithm when $1-(LB/UB)\leq \epsilon$, where $\epsilon$ is a predecided tolerance and/or, when consensus is reached.
\end{itemize}
This algorithm is an asynchronous counterpart of a previously presented algorithm in \cite{papavasiliou2013multiarea}, \cite{papavasiliou2015applying}, and \cite{papavasiliou2011reserve}. 
In all these previous references, the decomposition has been carried out across different scenarios among the shared variables (which happened to be the decision variables for switching states of slowly responding generators in the context of solving Stochastic Unit Commitment problem). Eventually an attempt is made to reach a consensus regarding the optimal values. Block Coordinate Descent (BCD) is an approximate method, provided that the error in the subgradient is bounded, and the stepsize in the algorithm is diminishing \cite{aravena2017distributed}. In this situation, the algorithm converges to an approximate solution of its dual problem if the original problem is convex (and if not, the dual gives a lower bound). Thus, in the proposed application, it is expected that when the algorithm converges and the error is bounded, it leads to the approximate optimum.
\subsection{Mathematical Model of Stage II Mechanism Design}\label{APP_mech}
\begin{figure}
\begin{center}
\includegraphics[scale=0.85]{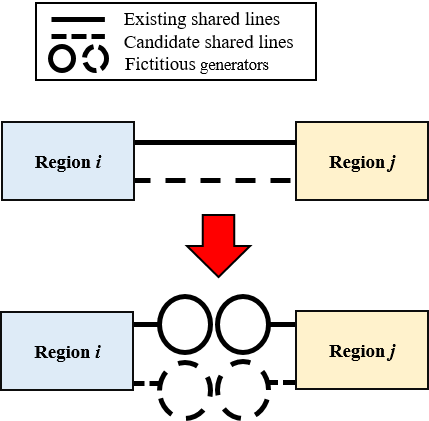}
\caption{Illustration of the APP layer of the incentive mechanism: centralized problem}
\label{APPMech_First}
\vspace{-5mm}
\end{center}
\end{figure}

The second stage of the mechanism is based on ``Auxiliary Problem Principle (APP)," which decides the generation values and the tie-line flows more accurately (if the accuracy in the first stage for these is not high enough). 
Let us first focus on the problem instances (\ref{second_area_decomposition}) which are the area decomposed versions of (\ref{Step2:General_one_Area}) with some modifications. 
In order to consider the inter-region transmission expansion, or the horizontal coordination as well as the existing lines shared between two different regions, it is necessary to calculate the flows on these existing and potential shared lines.
In the previous section the algorithm presented helps us determine which candidate lines will actually be considered for constructing new transmission lines as well as determining the first-order approximation of flows on both the existing lines as well as the ``to-be-constructed lines." 
Once we have reached a consensus regarding the optimal values of the discrete decision variables in the second stage, we  explicitly solve for the continuous variables. 
Our goal here is to come up with an incentive mechanism design such that the individual TPs are incentivized to act in a way that reaches the global optimum of the above-mentioned problem. 
In order to achieve this end it is important that each TP reach consensus about the flows on the shared lines. 
This is achieved with the \emph{Auxiliary Problem Principle (APP)} layer of the incentive mechanism. 
We assume each of the shared lines belonging to each region is replaced by a fictitious generator as shown in Fig. \ref{APPMech_First}. 
In Fig. \ref{APPMech_First} we have shown part of a power system network which is split into two regions on the two sides of the solid and dashed lines. 
There are two shared transmission lines, one existing (the solid line) and one candidate line (the dashed one) shared between the two regions. These two shared lines are notionally replaced by four generators---two for each line. The dashed generators represent the candidate transmission line and the solid generators are the ones for the existing transmission line. When the original centralized optimization problem is decomposed across the different regions, the flows on each line are treated equivalently to the outputs of these fictitious generators. 
However, each TP has to have some belief about the value of the shared variables and also about what the other TP's beliefs are about those respective variables. The APP iterations attempt to reach a consensus on those beliefs. \\
Let us consider a line $h\in \Tilde{H}$ shared between two regions, $z_1$ and $z_2$. The beliefs regarding the values of the node voltage angles at the two ends of the lines are as follows:
\begin{itemize}
    \item $\theta_{z_1,i(h)}$: Belief of $z_1$ about the voltage phase angle of the node belonging to itself
    \item $\theta_{z_1,j(h)}$: Belief of $z_1$ about the voltage phase angle of the node belonging to $z_2$
    \item $\theta_{z_2,i(h)}$: Belief of $z_2$ about the voltage phase angle of the node belonging to itself
    \item $\theta_{z_2,j(h)}$: Belief of $z_2$ about the voltage phase angle of the node belonging to $z_1$
\end{itemize}
Ideally $\theta_{z_1,i(h)}=\theta_{z_2,j(h)}$ and $\theta_{z_2,i(h)}=\theta_{z_1,j(h)}$ must be satisfied $\forall h\in \Tilde{H}$. 
However, this need not necessarily be true at the end of the algorithmic incentive mechanism computations introduced in Section \ref{Distributed}. 
The optimization problem formulations stated below describes the \emph{APP} iterations to achieve the consensus between the beliefs regarding the voltage phase angles and consequently the flows on the shared lines.
\begin{subequations}\label{APP}
\begin{gather}
\min_{P_{sg},\theta}\Bigg(\sum_{s\in S}\sum_{g\in{G_z}}w_sC_{g}(P_{sg})\notag\\+\sum_{z'}\sum_{h\in \Tilde{H}_z}\Big\{\eta\Big(\theta_{z,i(h)}(\theta^{\sigma}_{z,i(h)}-\theta^{\sigma}_{z',j(h)})\notag\\+\theta_{z,j(h)}(\theta^{\sigma}_{z,j(h)}-\theta^{\sigma}_{z',i(h)})\Big)+\frac{\gamma}{2}\Big(||\theta_{z,i(h)}-\theta^{\sigma}_{z,i(h)}||^2_2\notag\\+||\theta_{z,j(h)}-\theta^{\sigma}_{z,j(h)}||^2_2\Big)+\lambda^{\sigma}_{zz'i(h)}\theta_{z,i(h)}+\lambda^{\sigma}_{zz'j(h)}\theta_{z,j(h)}\Big\}\Bigg)\label{first_APP}\\
\mbox{Subject\;to:\:}\sum_{g\in G_{zn}}P_{sg}-\sum_{d\in D_{zn}}P_{sd}=\notag\\\sum_{h\in H_{zn}}\mathbf{A}_{(n,h)}P_{sh}+\sum_{h\in{\Tilde{H}_{zn}}}P_{sh}\notag\\\forall{z\in Z},\forall{s\in S},\forall{{n}\in\mathcal{N}_z}\label{second_APP}\\
P_{sh}=\frac{1}{x_h}\sum_{n\in \mathcal{N}}\mathbf{A}_{(n,h)}\theta_{sn},\forall{z\in Z},\forall{s\in S},\forall{{h}\in H_z}\label{third_APP}\\
P_{sh}=\frac{1}{x_h}(\theta_{szi(h)}-\theta_{szj(h)})\notag\\h \leftarrow (i,j),\forall{z\in Z},\forall{s\in S},\forall{{h}\in \Tilde{H}_z}\label{fourth_APP}\\
-{\overline{L}}_{h}\leq P_{sh}\leq{{\overline{L}}_{h}},\;\forall{z\in Z},\forall{s\in S},\forall{{h}\in H_z}\notag\\\label{fifth_Step2:General}\\
-{\overline{L}}_{h}\leq {P}_{sh}\leq {\overline{L}}_{h},\;\forall{z\in Z},\forall{s\in S},\forall{{h}\in \Tilde{H}_z}\notag\\\label{sixth1_APP}\\
{\underline{P}}_{sg}\leq P_{sg}\leq{{\overline{P}}_{sg}},\;\forall{z\in Z},\forall{s\in S},\forall{g\in{G_z}}\label{sixth_APP}\\
\text{Dual Variable Updates}\notag\\
\lambda^{\sigma+1}_{zz'i(h)}=\lambda^{\sigma}_{zz'i(h)}+\delta_{APP}(\theta^{\sigma+1}_{z,i(h)}-\theta^{\sigma+1}_{z',j(h)})\;\forall h\in \Tilde{H}\label{seventh_APP}\\
\lambda^{\sigma+1}_{zz'j(h)}=\lambda^{\sigma}_{zz'j(h)}+\delta_{APP}(\theta^{\sigma+1}_{z,j(h)}-\theta^{\sigma+1}_{z',i(h)})\;\forall h\in \Tilde{H}\label{eightth_APP}
\end{gather}
\end{subequations}
\subsection{Steps of Stage II Mechanism Design}\label{second_steps}
In the incentive mechanism design we assume the presence of a TPC as before. For example, in Europe the TPC is ENTSO-E. In North America, the closest entity to a TPC is FERC. 
This entity sets the penalty and designs the market rules, in order to incentivize the individual owners of transmission sub-network, to act in a way that solves the problem (\ref{APP}). 
In the following section we will explicitly state the steps or the rules of the second stage of the incentive mechanism design. 
In this stage each TP solves for the continuous variables more precisely. The steps of this stage of the mechanism can be described as follows:
\begin{itemize}
    \item Each region $z \in Z$ receives the values of the previous beliefs of voltage phase angles of the other node of the shared lines belonging to the adjoining regions, from the TPC. Regions $z$ also receive the previous updates of the rewards/penalties, which are the Lagrange Multiplier or dual variable values ($\theta^{\sigma}_{z',i(h)},\theta^{\sigma}_{z',j(h)}$ in (\ref{APP})).
    \item Each region independently solves the optimization problem (\ref{first_APP})--(\ref{sixth_APP}) and broadcasts the optimal decision variable values of its own voltage phase angles of the nodes of the shared lines to the TPC.
    \item The TPC calculates the rewards/penalties (Lagrange multipliers) by solving constraints (\ref{seventh_APP})--(\ref{eightth_APP}) and sends the updated values to the respective regions.
    \item The TPC stops the process when $(\theta^{\sigma+1}_{z,i(h)}-\theta^{\sigma+1}_{z',j(h)})^2+(\theta^{\sigma+1}_{z,j(h)}-\theta^{\sigma+1}_{z',i(h)})^2\leq \epsilon_{APP}$ for all the pairs of regions and all the shared lines, where $\epsilon_{APP}$ is the predetermined APP tolerance.
\end{itemize}
\subsection{The Proposed Organizational Set-up}\label{second_steps_sub}

In summary, this paper suggests the following set-up:
    \begin{itemize}
         \item \textbf{Step 1:} the TPC announces the incentive mechanism for stage I and asks the TPs to submit their candidate lines for inter- and intra-regional planning. 
        \item \textbf{Step 2:} the TPC solves its optimization problem in stage I and sends the penalty and rewards to TPs. 
        \item \textbf{Step 3:} TPs correct their decisions based on the penalty and rewards they have received. The new decisions are sent to TPC 
        \item \textbf{Step 4:} the TPC iterates between Steps 2 and 3 to reach a consensus. 
        \item \textbf{Step 5:} Once the consensus is reached the inter- and intra-regional planning results will be published. The TPC announces the incentive mechanism for stage II and the same steps as stage I are run for the second stage. Finally, with the final results, TPs can start constructing the planned lines. 
        \end{itemize}
\section{Numerical Results}\label{numerical}
\subsection{2-region System}
\begin{figure}
\begin{center}
\includegraphics[scale=0.6]{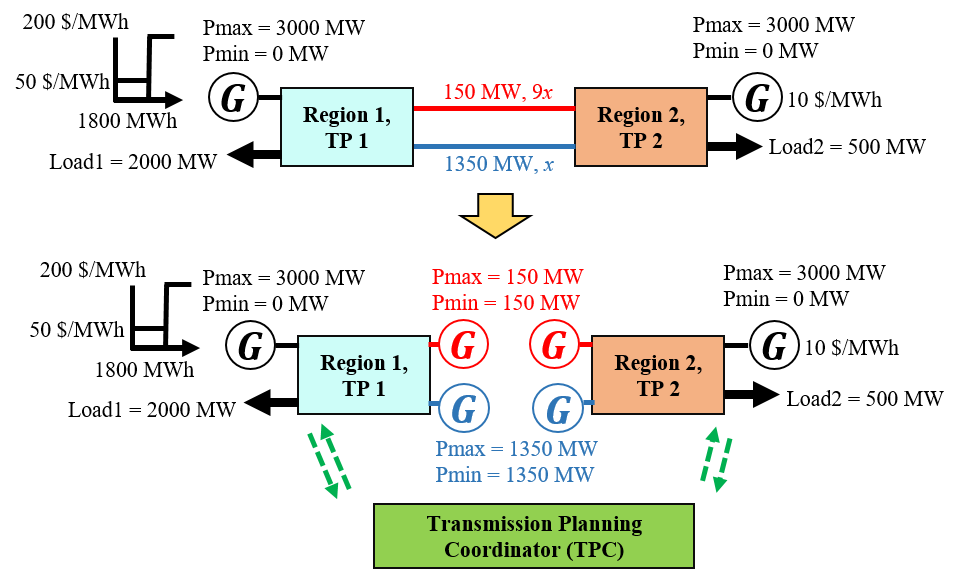}
\caption{Two-region power system for comparison between game-theoretic and mechanism design results}
\label{Mech2Bus}
\end{center}
\end{figure}

In this sub-section, an illustrative two-region power system is considered to present our novel mechanism design compared to a game-theoretic design. For the system shown in Figure~\ref{Mech2Bus}, Tables \ref{table:2RegionSystem} and \ref{table:2RegionSharedLines} provide the system details.
\begin{table}[ht] 

\caption{The data of the two-region system} 

\centering 

\begin{tabular}{ c  c  c } 

\hline\hline 

Parameter & Region-1 & Region-2\\ [0.5ex] 


\hline 


Gen. Range (MW) & 0-3000 & 0-3000 \\ 
\hline 
Gen. Cost (\$/MWh) & 50 for 0-1800 MW & 10 \\  
   & 200 for 1800-300 MW &   \\   
\hline 
Load (MW) & 2000 & 500 \\   
\hline 
First-case Investment (\$) & 1000 & 1000 \\  
\hline 
second-case Investment (\$) & 20000 & 20000 \\   
\hline 
\end{tabular} 

\label{table:2RegionSystem} 
\end{table} 
\begin{table}[ht] 

\caption{Shared Lines of two-Region System} 

\centering 

\begin{tabular}{ c  c  c } 

\hline\hline 

Parameter & Existing & Candidate\\ [0.5ex] 


\hline 


Capacity (MW) & 150 & 1350 \\ 
\hline 
Reactance (pu) & 9X & X \\  [1ex] 
\hline 
\end{tabular} 

\label{table:2RegionSharedLines} 
\end{table}
The line along the candidate line is constructed \textit{iff} both TP-1 and TP-2 think that the line should be constructed. 
We have considered here two cases. In the first case, the total investment cost is \$2000 and is shared equally between TP-1 and TP-2 if the line is constructed. 
\begin{figure}
\begin{center}
\includegraphics[scale = 0.85,clip]{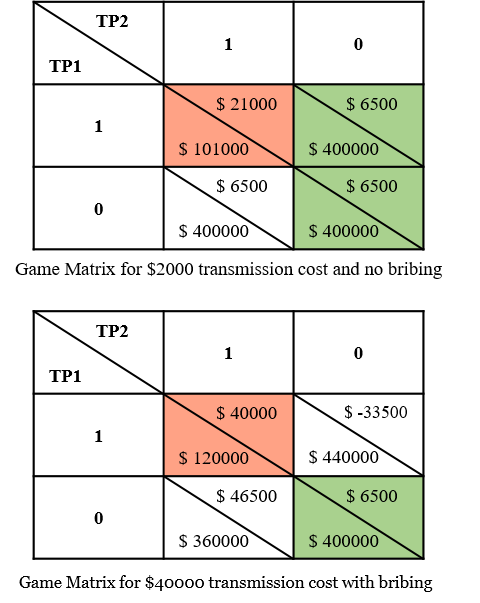}
\caption{Game matrix for two-region power system}
\label{gamematrix}
\vspace{-5mm}
\end{center}
\end{figure}
The two-player cost matrix for the game of strategic interaction between TP-1 and TP-2 in this case is shown at the top of Fig.~\ref{gamematrix}. 
The numbers shown are the costs faced by the individual regions. The numbers 1 and 0 represent the decisions to build and to not build the transmission line by each of the players, respectively. 
The social cost for the (1,1) entry is \$47000, whereas for each of the other three boxes is \$378500. 
Hence, the box that is in orange (corresponding to the transmission line being built) is clearly the social optimum (SO, or Pareto optimum). 
The game-theoretic Nash Equilibrium (NE) are the boxes in green, and we can see that there are two of them. 

In the second case, we consider the situation where the cost for building the new line is \$40000 which is shared equally among the two players. 
If one player decides to build the line when another one does not, then the player who decides to build the line bribes the other one in the hope of changing their mind in the future. 
The cost matrix for such a game is shown at the bottom half of Fig.~\ref{gamematrix}. 
Here, the SO is still the box (1,1), whereas now there is only one NE in box (0,0). 
In both the cases examined, the SO and NE are not the same strategy combination. 
This is a serious drawback of the game-theoretic interaction, which our mechanism attempts to resolve. 

In order to show how our mechanism can help resolve the diffrence between SO and NE, we now refer to the modified model of the system shown in the bottom half of Fig.~\ref{Mech2Bus}. 
In this diagram, we have introduced the TPC with the green lines representing the message exchange pertaining to the rewards/penalties and tentative decisions for building the lines. 
The shared lines have been split up and represented as virtual generators, with the line capacities for forward and reverse flows acting as generating limits and the investment cost can be interpreted as the start-up cost of these fictitious generators. 
The mechanism is implemented in different steps or iterations. 
Before the start of each round or iteration of the mechanism, the TPC broadcasts the most recent estimate of the rewards or penalties ($\pi, \mu_k, \mu_h$ respectively refer to the rewards/penalties associated with the non-anticipativity constraints for consensus for decision to build transmission line, flow on the candidate line, and flow on the existing shared line). 
Each region then solves its own optimization problem (calculating $u^*, P^*_k, P^*_h, P^*_g$, which respectively refer to the integer variable for decision to build transmission line, flow on the candidate line, flow on the existing shared line, and the generator output) and relays its tentative decision to the TPC, following which the TPC updates the rewards/penalties. This goes on until consensus is reached. At the beginning of the first iteration, all the rewards/penalties are set to zero. The details of each step are shown in Table~\ref{table:DispatchIntervalConvention}:
\begin{table}[ht] 

\caption{Iterates of the incentive mechanism} 

\centering 

\begin{tabular}{c  c  c  c  c } 

\hline\hline 

Iterates & Iter-1 & Iter-2 & Iter-3 & Iter-4 \\ [0.5ex] 


\hline 


$u^{TP1*}$ & 1 & 1 & 1 & 1 \\ 
\hline 
$P^{TP1*}_k$ & 1350 & 180 & 180 & 1350 \\  [1ex] 
\hline 
$P^{TP1*}_h$ & 150 & 20 & 20 & 150 \\  [1ex] 
\hline 
$P^{TP1*}_g$ & 500 & 1800 & 1800 & 500 \\  [1ex] 
\hline 
$u^{TP2*}$  & 0 & 0 & 1 & 1 \\ 
\hline 
$P^{TP2*}_k$ & 0 & 0 & -1350 & -1350 \\  [1ex] 
\hline 
$P^{TP2*}_h$ & 0 & 0 & -150 & -150 \\  [1ex] 
\hline 
$P^{TP2*}_g$ & 500 & 500 & 2000 & 2000 \\  [1ex] 
\hline 
$u^{MO*}$ & 0 & 1 & 1 & 1 \\  [1ex] 
\hline 
$P^{MO*}_k$ & 0 & 1350 & 1350 & 1350 \\  [1ex] 
\hline 
$P^{MO*}_h$ & 0 & 150 & 150 & 150 \\  [1ex] 
\hline 
$\pi^{TP1*}$ & 1 & 1 & 1 & 1 \\  [1ex] 
\hline 
$\mu^{TP1*}_k$ & 1350 & 180 & -990 & -990 \\  [1ex] 
\hline 
$\mu^{TP1*}_h$ & 150 & 20 & -110 & -110 \\  [1ex] 
\hline 
$\pi^{TP2*}$ & 0 & -1 & -1 & -1 \\  [1ex] 
\hline 
$\mu^{TP2*}_k$ & 0 & 1350 & 1350 & 1350 \\  [1ex] 
\hline 
$\mu^{TP2*}_h$ & 0 & 150 & 150 & 150 \\  [1ex] 
\hline 
$(1-UB^*/LB^*)$ & -0.01 & 15.78 & 0.0002 & 0 \\  [1ex] 
\hline 
\end{tabular} 

\label{table:DispatchIntervalConvention} 
\end{table} 
As we can see from Table~\ref{table:DispatchIntervalConvention}, indeed our proposed mechanism reaches the social optimum and the finally settled values of the rewards/penalties are precisely the incentivizing factors needed to force the behavior of each TP away from the NE and towards the SO. 
\subsection{Generalized/Multi-region System}
\begin{figure}
\begin{center}
\includegraphics[scale=0.55]{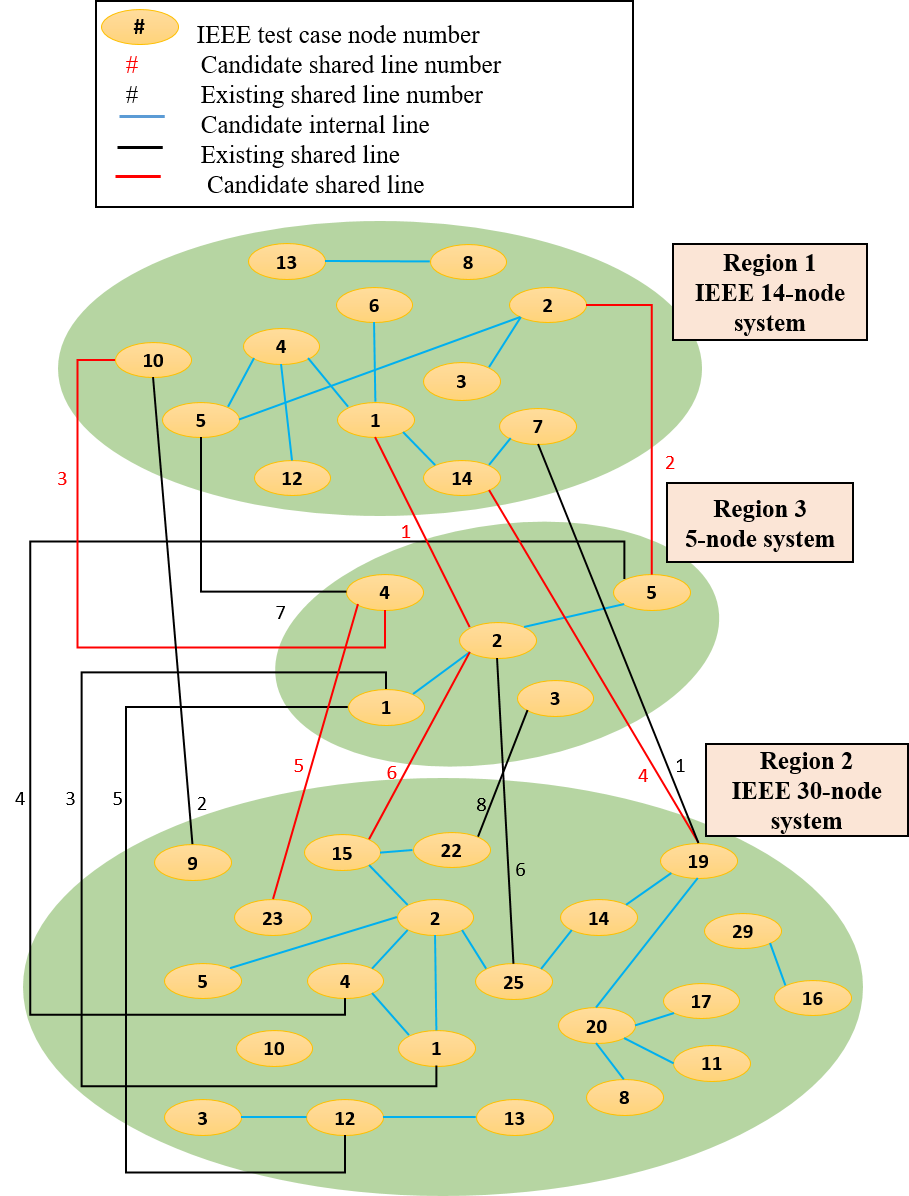}
\caption{The multi-regional system considered for the simulation}
\label{simSys}
\vspace{-5mm}
\end{center}
\end{figure}

In Fig. ~\ref{simSys}, we have considered a bigger system, which consists of three regions. Regions 1, 2, and 3, respectively are the IEEE 14-, 30-, and 5-node systems. The red lines are the inter-regional shared candidate lines, the blue lines represent the intra-regional candidate lines, and the black ones are the existing shared lines.

\begin{figure}
\begin{center}
\includegraphics[scale = 0.145]{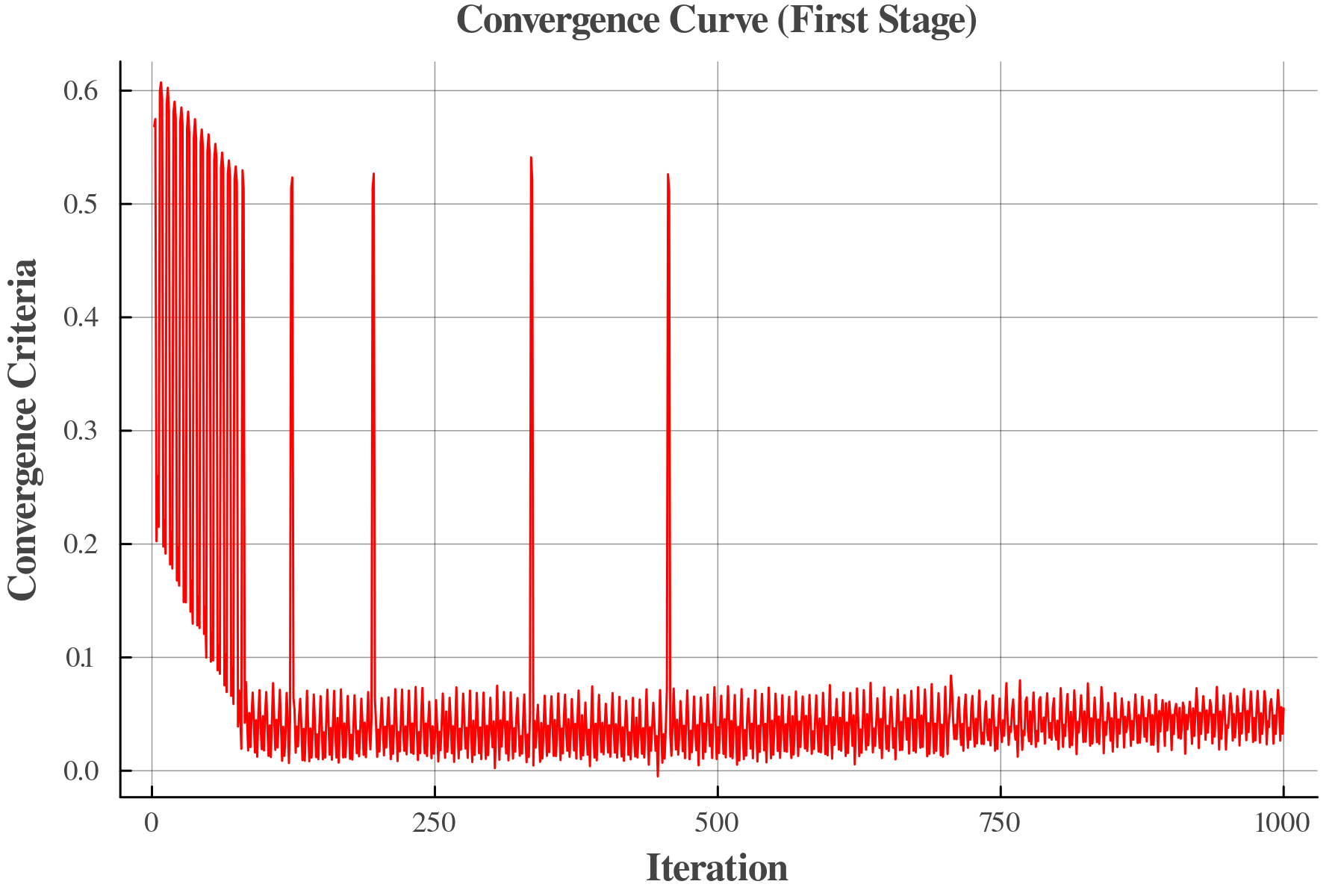}
\caption{Variation of convergence criteria for the first stage according to iterations}
\label{stage1}
\vspace{-5mm}
\end{center}
\end{figure}
We apply the stage-I incentive mechanism design algorithm for 1000 iterations in order to illustrate the trend and variations of the convergence curve. We get the convergence at the 85th iteration as illustrated in Fig. ~\ref{stage1}. The convergence curve indicates the difference between the low and high bounds of the distributed problem. Fig. ~\ref{stage1} explains that the convergence curve gradually reaches a stable value after 500 iterations. The big spikes in the convergence curve, for example, the ones happening at iterations 123, 195, 335, and 456 indicate the discrepancy between the binary decision of building the fourth line. They state that the decision of the TPC does not match those of TPs. The smaller oscillations refer to a minor disagreement in the operational decisions. They can be disregarded since their domain is relatively small and the aim of the first stage is to determine binary decision variables related to the candidate lines' constructions.

\begin{figure}
\begin{center}
\includegraphics[scale=0.145,clip]{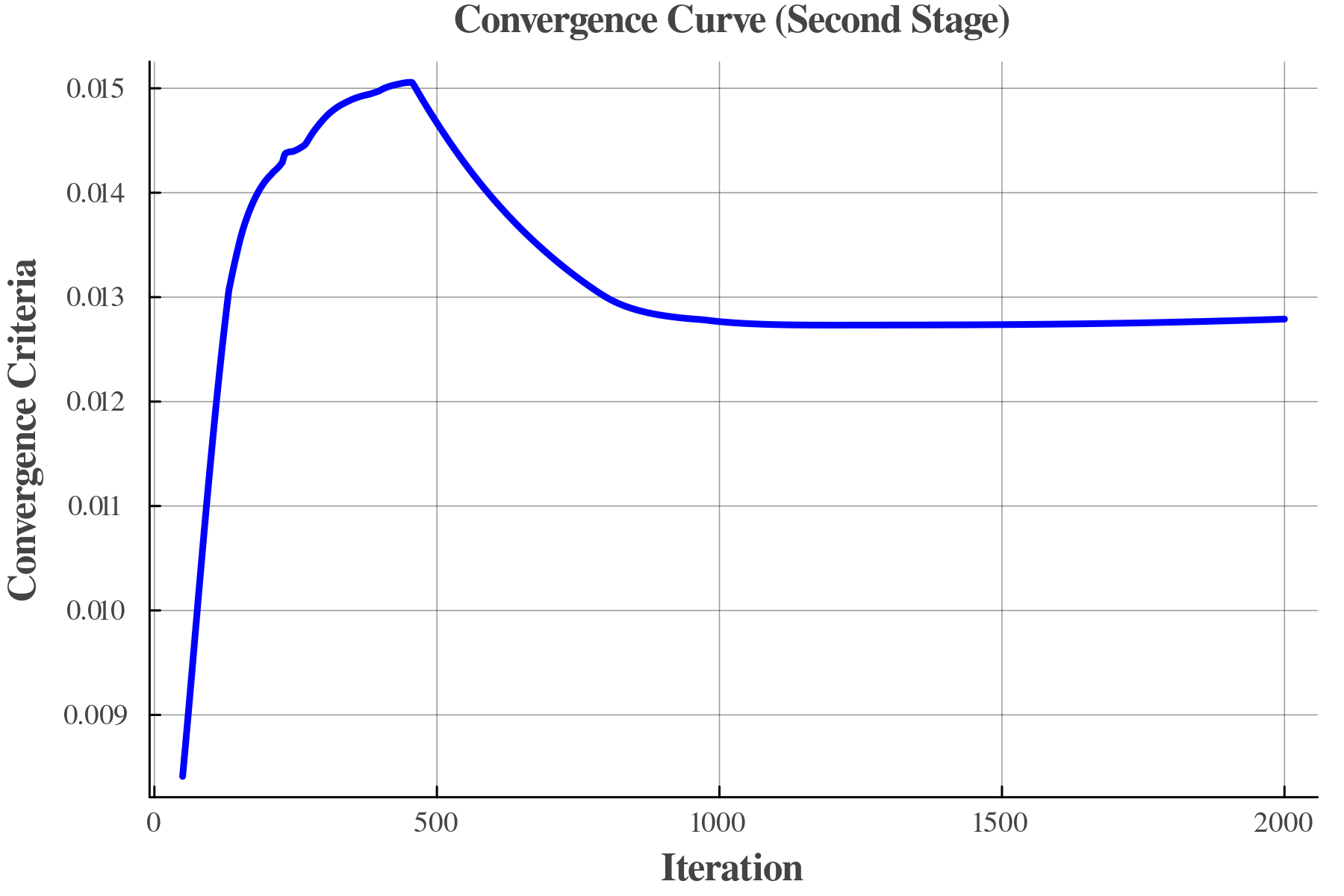}
\caption{Variation of convergence criteria for the second stage according to iterations}
\label{stage2}
\vspace{-5mm}
\end{center}
\end{figure}
According to the given system, Region 2 has a high load, and expensive generation whereas Regions 1 and 3 have low and moderate loads with cheap and moderately cheap generations, respectively. 
We have observed that in this situation, the optimal decision is to build candidate line 4 which is shared between Regions 1 and 2. \\
In addition, Fig.~\ref{stage2} shows the variation of the convergence curve for the second stage. We have initialized variables to zeros. Hence the convergence starts with a small value. After almost 1000 iterations, all TPs reach an agreement on the operational variables. The obtained convergence criterion is equal to 0.012.\\
Finally, Table~\ref{table:Intra-Construc} compares the intra-regional candidate decisions considering centralized and the proposed distributed models. Correspondingly, Table~\ref{table:Inter-Construc} compares the inter-regional candidate decisions when the results are determined by the centralized and the proposed distributed methods. \\
We should note that the centralized framework is just a mathematical benchmark in our paper and it does not exist in the real-world situations. In real-life situations, some level of coordination always exists. In our paper, we suggest an optimization-based coordination for the investment coordination problem. We have shown that our mechanism can reach a close-to-global solution as compared to our theoretical benchmark model. At our close-to-optimal solution, each TP receives a share of total social welfare which is optimal for the whole inter-connected system. Hence, it will achieve a solution close to the original optimal solution. The error is mainly due to the non-convex nature of the whole optimization problem.\\
We have run this  simulation on a desktop computer using Julia programming language with JuMP optimization interface and with several different solvers (for the first stage we have used MILP solvers such as Gurobi, HiGHS, GLPK, and for the second stage which is a Quadratic Program (QP) we have used solvers such as Gurobi and IPOPT). 
Codes are available to be run and verified at \url{http://github.com/sambuddhac/Horizontal_Proper}.
\begin{table}[ht] 

\caption{Intra-regional Candidate Line Construction Decisions} 

\centering 

\begin{tabular}{ c  c  c  c } 

\hline\hline 

Mechanism & Region-1 & Region-2 & Region-3 \\ [0.5ex] 


\hline 


Centralized & 10 & 1,10,11,14,16 & -- \\ 
\hline 
Dist. Mech. & -- & -- & -- \\  [1ex] 
\hline 
\end{tabular} 

\label{table:Intra-Construc} 
\end{table} 
\begin{table}[ht] 

\caption{Inter-regional Candidate Line Construction Decisions} 

\centering 

\begin{tabular}{ c  c  c  c  c  c } 

\hline\hline 

Mechanism & Region-1 & Region-2 & Region-3 & TPC & Decision\\ [0.5ex] 


\hline 


Centralized & 4 & 4,5 & 5 & 4,5 & 4,5 \\ 
\hline 
Dist. Mech. & 4 & 4 & -- & 4 & 4 \\  [1ex] 
\hline 
\end{tabular} 

\label{table:Inter-Construc} 
\end{table} 
\section{Conclusion}\label{conclusion}
In this paper, we have addressed the important problem of transmission investment when there are multiple regions and new lines built between them. 
We have proposed a two-staged incentive mechanism design consisting of a region-decomposed distributed Lagrangian method. 
In our numerical study, we have analyzed an illustrative two-region network and a larger and more general three-region system. The two-region network results intuitively analyzed and validated our incentive mechanism approach, which includes a TPC, compared to the game-theoretic Nash equilibrium approach that does not have a TPC \cite{tohidi2013multi,tohidi2014multi,tohidi2017coordination}. 
We showed that our proposed incentive mechanism indeed reached the Pareto optimal solution which is also the socially optimal solution. For the three-region case, we implemented our proposed mechanism design in a computational framework to simulate the mechanism performance. We presented the convergence characteristics of both the stages of our mechanism and discussed the results. Future research steps include: \begin{itemize}
\item integrating our model with generation expansion and consider policies such as capacity reserve margin, renewable portfolio standard and emissions margin.
\item including $(N-1)$ contingency criteria within our transmission expansion planning problem.
\item including temporal variation, inter-temporal constraints, and time-domain reduction.
\end{itemize}
As a limitation of implementing the proposed algorithms, one challenge would
be related to defining the appropriate stepsize in the first and second
algorithms in order to help the convergence of the algorithm and avoid
unbounded error as well as a compromise between the computational costs and
accuracy of the final solutions. This issue can be further analyzed and
resolved in future research.

\printbibliography 

@mastersthesis{Sambuddha2010,
    title    = {Study of UPLAN based Resources Planning \& Analysis by Power Generation Utilities in the Deregulated Electricity Market},
    school   = {The University of Texas at Austin},
    author   = {Chakrabarti, S},
    year     = {2010}, %other attributes omitted
}

@article{wang2017survey,
  title={A survey on energy internet: Architecture, approach, and emerging technologies},
  author={Wang, Kun and Yu, Jun and Yu, Yan and Qian, Yirou and Zeng, Deze and Guo, Song and Xiang, Yong and Wu, Jinsong},
  journal={IEEE systems journal},
  volume={12},
  number={3},
  pages={2403--2416},
  year={2017},
  publisher={IEEE}
}

@article{hou2021incentive,
  title={Incentive-driven task allocation for collaborative edge computing in industrial internet of things},
  author={Hou, Wenjing and Wen, Hong and Zhang, Ning and Wu, Jinsong and Lei, Wenxin and Zhao, Runhui},
  journal={IEEE Internet of Things Journal},
  volume={9},
  number={1},
  pages={706--718},
  year={2021},
  publisher={IEEE}
}

@article{zeng2021novel,
  title={A novel reputation incentive mechanism and game theory analysis for service caching in software-defined vehicle edge computing},
  author={Zeng, Feng and Chen, Yaojia and Yao, Lan and Wu, Jinsong},
  journal={Peer-to-Peer Networking and Applications},
  volume={14},
  pages={467--481},
  year={2021},
  publisher={Springer}
}

@article{eia2014electricity,
  title={The Electricity Market Module of the National Energy Modeling System: Model Documentation 2020},
  author={EIA, US},
  journal={US Energy Information Administration, Washington, DC, Tech. Rep},
  year={2020}
}

@misc{epa2018documentation,
  title={Documentation for EPA’s Power Sector Modeling Platform v6-November 2018 Reference Case},
  author={EPA},
  year={2018},
  publisher={US Environmental Protection Agency Washington, DC}
}

@techreport{young2018us,
  title={US-REGEN Model Documentation},
  author={Young, D and Blanford, G and Bistline, J and Rose, S and de la Chesnaye, F and Bedilion, R and Wilson, T and Wan, S},
  year={2018},
  institution={Technical report, Electric Power Research Institute, Palo Alto, CA}
}

@techreport{ho2021regional,
  title={Regional Energy Deployment System (ReEDS) Model Documentation: Version 2020},
  author={Ho, Jonathan and Becker, Jonathon and Brown, Maxwell and Brown, Patrick and Chernyakhovskiy, Ilya and Cohen, Stuart and Cole, Wesley and Corcoran, Sean and Eurek, Kelly and Frazier, Will and others},
  year={2021},
  institution={National Renewable Energy Lab.(NREL), Golden, CO (United States)}
}

@techreport{mai2013resource,
  title={Resource Planning Model: An Integrated Resource Planning and Dispatch Tool for Regional Electric Systems},
  author={Mai, Trieu and Drury, Easan and Eurek, Kelly and Bodington, Natalie and Lopez, Anthony and Perry, Andrew},
  year={2013},
  institution={National Renewable Energy Lab.(NREL), Golden, CO (United States)}
}

@article{johnston2019switch,
  title={Switch 2.0: A Modern Platform for Planning High-Renewable Power Systems},
  author={Johnston, Josiah and Henriquez-Auba, Rodrigo and Maluenda, Benjam{\'\i}n and Fripp, Matthias},
  journal={SoftwareX},
  volume={10},
  pages={100251},
  year={2019},
  publisher={Elsevier}
}

@article{brown2017pypsa,
  title={PyPSA: Python for Power System Analysis},
  author={Brown, Tom and H{\"o}rsch, Jonas and Schlachtberger, David},
  journal={arXiv preprint arXiv:1707.09913},
  year={2017}
}

@article{jenkins2017enhanced,
  title={Enhanced Decision Support for a Changing Electricity Landscape: The GenX Configurable Electricity Resource Capacity Expansion Model},
  author={Jenkins, Jesse D and Sepulveda, Nestor A},
  year={2017},
  publisher={MIT Energy Initiative}
}

@phdthesis{jenkins2018electricity,
  title={Electricity System Planning with Distributed Energy Resources: New Methods and Insights for Economics, Regulation, and Policy},
  author={Jenkins, Jesse David},
  year={2018},
  school={Massachusetts Institute of Technology}
}

@article{tohidi2014multi,
  title={Multi-Regional Transmission Planning as a Non-Cooperative Decision-Making},
  author={Tohidi, Y and Hesamzadeh, M R},
  journal={IEEE Transactions on Power Systems},
  volume={29},
  number={6},
  pages={2662--2671},
  year={2014},
  publisher={IEEE}
}

@article{tohidi2016optimal,
  title={Optimal Long-Term Generation-Transmission Planning in the Context of Multiple TSOs},
  author={Tohidi, Y},
  year={2016}
}

@inproceedings{tohidi2013free,
  title={Free Riding Effect in Multi-National Transmission Expansion Planning},
  author={Tohidi, Y and Hesamzadeh, M R},
  booktitle={Innovative Smart Grid Technologies Europe (ISGT EUROPE), 2013 4th IEEE/PES},
  pages={1--5},
  year={2013},
  organization={IEEE}
}

@inproceedings{tohidi2013multi,
  title={Multi-National Transmission Planning Using Joint and Disjoint Solutions},
  author={Tohidi, Y and Hesamzadeh, M R},
  booktitle={European Energy Market (EEM), 2013 10th International Conference on the},
  pages={1--6},
  year={2013},
  organization={IEEE}
}

@inproceedings{tohidi2014multiwind,
  title={Multi-Regional Transmission Planning Under Interdependent Wind Uncertainty},
  author={Tohidi, Y and Hesamzadeh, M R},
  booktitle={Energy Conference (ENERGYCON), 2014 IEEE International},
  pages={1474--1479},
  year={2014},
  organization={IEEE}
}

@article{tohidi2017coordination,
  title={Coordination of Generation and Transmission Development Through Generation Transmission Charges—A Game Theoretical Approach},
  author={Tohidi, Y and Olmos, L and Rivier, M and Hesamzadeh, M R},
  journal={IEEE Transactions on Power Systems},
  volume={32},
  number={2},
  pages={1103--1114},
  year={2017},
  publisher={IEEE}
}

@inproceedings{tohidi2015reactive,
  title={Reactive Coordination of Transmission-Generation Investment Planning},
  author={Tohidi, Y and Hesamzadeh, M R and Ostman, K},
  booktitle={European Energy Market (EEM), 2015 12th International Conference on the},
  pages={1--5},
  year={2015},
  organization={IEEE}
}

@article{tohidi2017sequential,
  title={Sequential Coordination of Transmission Expansion Planning with Strategic Generation Investments},
  author={Tohidi, Y and Hesamzadeh, M R and Regairaz, F},
  journal={IEEE Transactions on Power Systems},
  volume={32},
  number={4},
  pages={2521--2534},
  year={2017},
  publisher={IEEE}
}

@article{aravena2017distributed,
	title={A Distributed Computing Architecture for the Large-Scale Integration of Renewable Energy and Distributed Resources in Smart Grids},
	author={Aravena, S I A and Papavasiliou, A and Papalexopoulos, A},
	year={2017}
}

@article{papavasiliou2011reserve,
  title={Reserve Requirements for Wind Power Integration: A Scenario-based Stochastic Programming Framework},
  author={Papavasiliou, A and Oren, S S and O'Neill, R P},
  journal={IEEE Transactions on Power Systems},
  volume={26},
  number={4},
  pages={2197--2206},
  year={2011},
  publisher={IEEE}
}

@article{papavasiliou2015applying,
  title={Applying High Performance Computing to Transmission-Constrained Stochastic Unit Commitment for Renewable Energy Integration},
  author={Papavasiliou, A and Oren, S S and Rountree, B},
  journal={IEEE Transactions on Power Systems},
  volume={30},
  number={3},
  pages={1109--1120},
  year={2015},
  publisher={IEEE}
}

@article{papavasiliou2013multiarea,
	title={Multiarea Stochastic Unit Commitment for High Wind Penetration in a Transmission Constrained Network},
	author={Papavasiliou, A and Oren, S S},
	journal={Operations Research},
	volume={61},
	number={3},
	pages={578--592},
	year={2013},
	publisher={INFORMS}
}

@phdthesis{Sambuddha2017,
    title    = {Post-Contingency States Representation \& Redispatch for Restoration in Power Systems Operation},
    school   = {The University of Texas at Austin},
    author   = {Chakrabarti, S},
    year     = {2017}, %other attributes omitted
}

@article{baldick1999fast,
  title={A Fast Distributed Implementation of Optimal Power Flow},
  author={Baldick, R and Kim, B H and Chase, C and Luo, Y},
  journal={IEEE Transactions on Power Systems},
  volume={14},
  number={3},
  pages={858--864},
  year={1999},
  publisher={IEEE}
}

@article{cohen1978optimization,
  title={Optimization by Decomposition and Coordination: A Unified Approach},
  author={Cohen, G},
  journal={IEEE Transactions on automatic control},
  volume={23},
  number={2},
  pages={222--232},
  year={1978},
  publisher={IEEE}
}

@article{cohen1980auxiliary,
  title={Auxiliary Problem Principle and Decomposition of Optimization Problems},
  author={Cohen, G},
  journal={Journal of optimization Theory and Applications},
  volume={32},
  number={3},
  pages={277--305},
  year={1980},
  publisher={Springer}
}

@article{kim1997coarse,
  title={Coarse-Grained Distributed Optimal Power Flow},
  author={Kim, B H and Baldick, R},
  journal={IEEE Transactions on Power Systems},
  volume={12},
  number={2},
  pages={932--939},
  year={1997},
  publisher={IEEE}
}

@article{kim2000comparison,
  title={A Comparison of Distributed Optimal Power Flow Algorithms},
  author={Kim, B H and Baldick, R},
  journal={IEEE Transactions on Power Systems},
  volume={15},
  number={2},
  pages={599--604},
  year={2000},
  publisher={IEEE}
}

@article{ebrahimian2000state,
  title={State Estimation Distributed Processing [for power systems]},
  author={Ebrahimian, R and Baldick, R},
  journal={IEEE Transactions on Power Systems},
  volume={15},
  number={4},
  pages={1240--1246},
  year={2000},
  publisher={IEEE}
}

@phdthesis{magnusson2017bandwidth,
  title={Bandwidth Limited Distributed Optimization with Applications to Networked Cyberphysical Systems},
  author={Magn{\'u}sson, S},
  year={2017},
  school={KTH Royal Institute of Technology}
}

@article{majidiStoch,
author = {Majidi-Qadikolai, Mohammad and Baldick, R.},
year = {2016},
month = {02},
pages = {1-12},
title = {Stochastic Transmission Capacity Expansion Planning With Special Scenario Selection for Integrating N-1 Contingency Analysis},
volume = {31},
journal = {IEEE Transactions on Power Systems},
doi = {10.1109/TPWRS.2016.2523998}
}

@article{majidiIntegration,
author = {Majidi-Qadikolai, Mohammad and Baldick, R.},
year = {2015},
month = {06},
pages = {1-12},
title = {Integration of $N-1$ Contingency Analysis With Systematic Transmission Capacity Expansion Planning: ERCOT Case Study},
volume = {31},
journal = {IEEE Transactions on Power Systems},
doi = {10.1109/TPWRS.2015.2443101}
}

@article{majidiDecomposition,
author = {Majidi-Qadikolai, Mohammad and Baldick, R.},
year = {2017},
month = {07},
pages = {1-1},
title = {A Generalized Decomposition Framework for Large-Scale Transmission Expansion Planning},
volume = {PP},
journal = {IEEE Transactions on Power Systems},
doi = {10.1109/TPWRS.2017.2724554}
}

@inproceedings{majidiReducing,
author = {Majidi-Qadikolai, Mohammad and Baldick, R.},
year = {2015},
month = {10},
pages = {1-6},
title = {Reducing the number of candidate lines for high level Transmission Capacity Expansion Planning under uncertainties},
doi = {10.1109/NAPS.2015.7335117}
}

@article{majidiRedSwitch,
author = {Majidi-Qadikolai, Mohammad and Baldick, R.},
year = {2015},
month = {07},
pages = {},
title = {Reducing the Candidate Line List for Practical Integration of Switching into Power System Operation}
}

@article{chakrabarti2020look,
  title={Look-Ahead SCOPF (LASCOPF) for Tracking Demand Variation via Auxiliary Proximal Message Passing (APMP) Algorithm},
  author={Chakrabarti, Sambuddha and Baldick, Ross},
  journal={International Journal of Electrical Power \& Energy Systems},
  volume={116},
  pages={105533},
  year={2020},
  publisher={Elsevier}
}

@article{chakrabarti2019look,
  title={Look-Ahead SCOPF (LASCOPF) for Tracking Demand Variation via Auxiliary Proximal Message Passing (APMP) Algorithm},
  author={Chakrabarti, Sambuddha and Baldick, Ross},
  journal={arXiv},
  volume={1},
  year={2019},
  publisher={arXiv}
}
\begin{IEEEbiography}[{\includegraphics[width=1in,height=1.25in,clip,keepaspectratio]{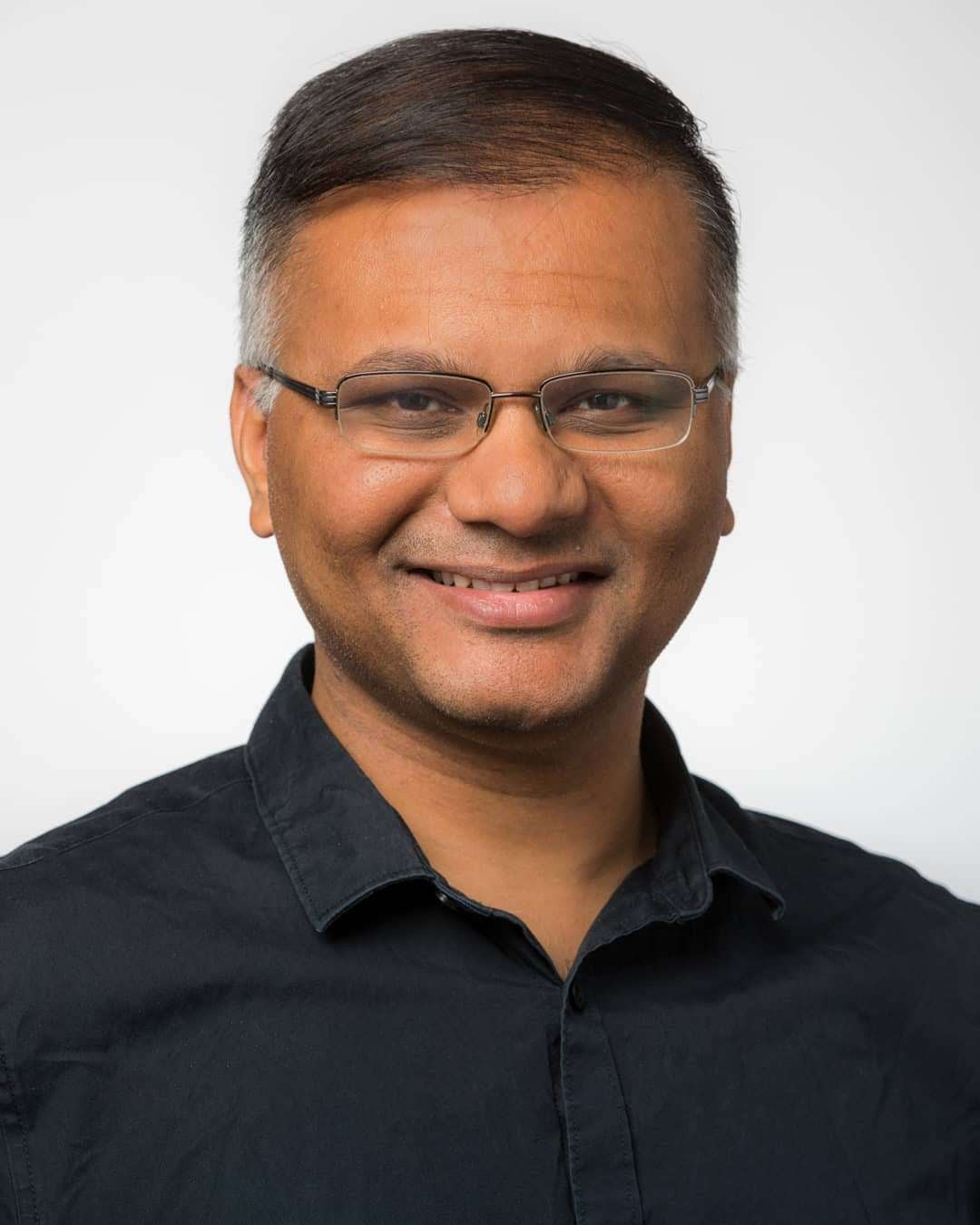}}]{Sambuddha Chakrabarti} (S'14-M'22) is an energy-systems researcher with the Andlinger Center for Energy + the Environment at Princeton University. From 2017 to 2020 he was a postdoctoral researcher at National Renewable Energy Laboratory (NREL), KTH, Linkoping University, Arizona State University, and  The University of Texas at Austin. He received his MS and Ph.D. from the Department of Electrical and Computer Engineering at The University of Texas at Austin. His research focuses on application of Operations Research methods to power network planning and operations problems
\end{IEEEbiography}

\begin{IEEEbiography}[{\includegraphics[width=1in,height=1.25in,clip,keepaspectratio]{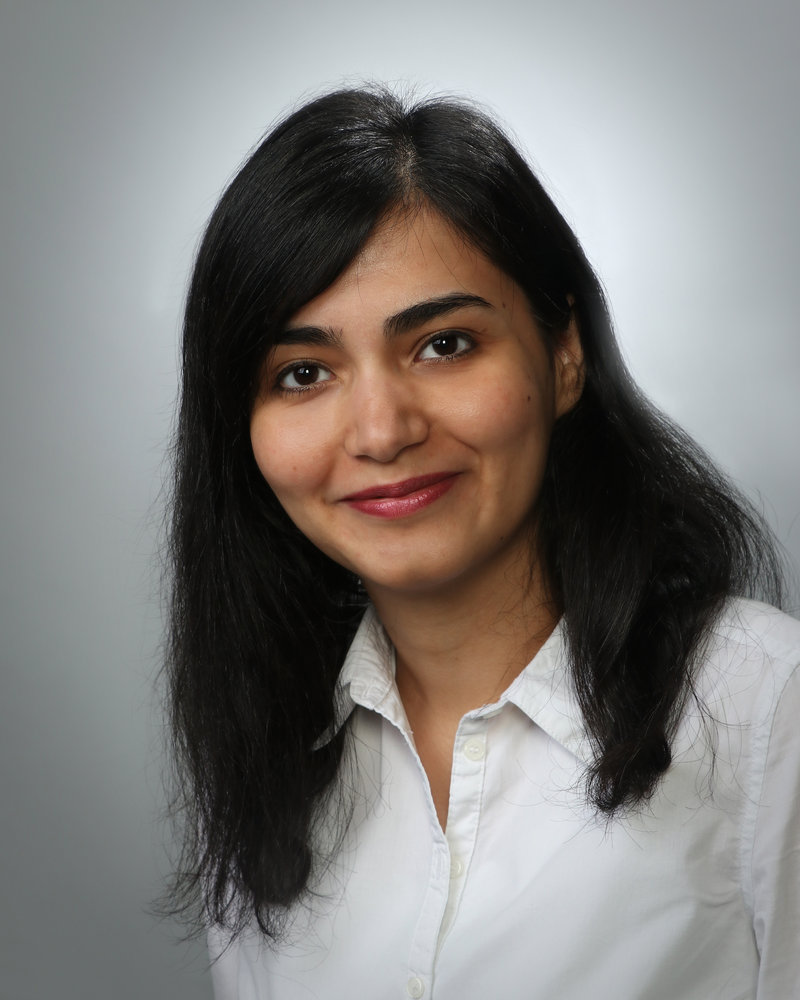}}]{Hosna Khajeh}(S'10)
 received the M.Sc. (Tech.) degree in electrical engineering (power systems) from Semnan University, Semnan, Iran in 2016. She is currently pursuing a doctoral degree at the University of Vaasa, Vaasa, Finland. Her research interests include electricity and ancillary service markets, energy flexibility management, forecasting, and trading structures.   
\end{IEEEbiography}


\begin{IEEEbiography}
[{\includegraphics[width=1in,height=1.25in,clip,keepaspectratio]{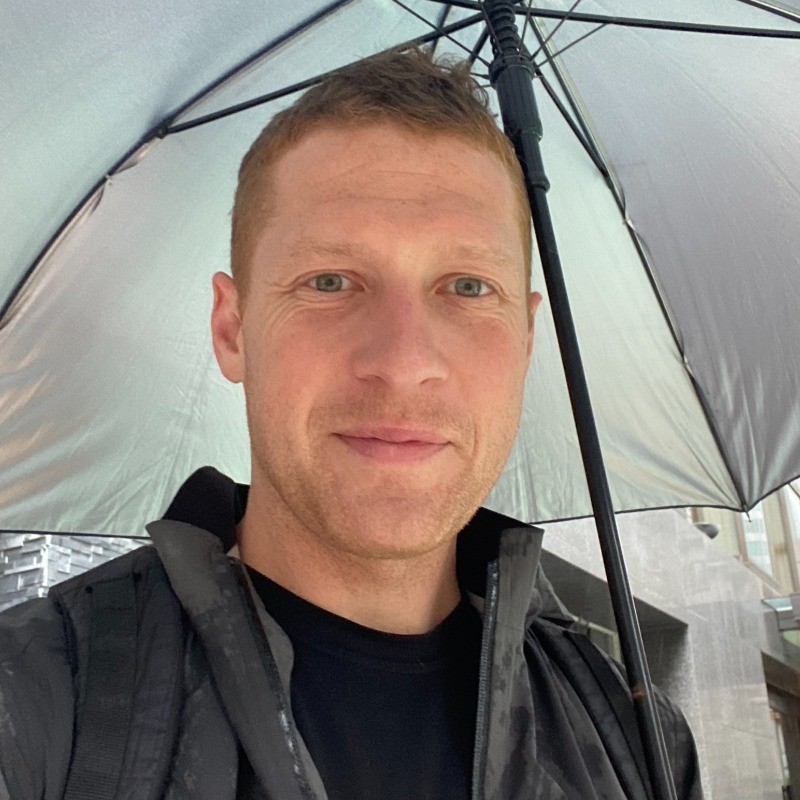}}]{Thomas R Nudell}
received the B.A. magna cum laude degree in physics from Kalamazoo College in 2009, the M.S. and Ph.D. degrees from North Carolina State University in 2011 and 2015, respectively, both in electrical engineering. He was then a postdoctoral research associate with the Active Adaptive Control Laboratory, Massachusetts Institute of Technology, where he investigated resilient interconnected infrastructures and transactive control for microgrid energy management. His professional goals include realizing the rapid and widespread integration of renewable and energy efficient resources in the electric power grid. Dr Nudell is the cofounder and CEO of Piq Energy Corp, building autonomous grid planning and operation software.
\end{IEEEbiography}

\begin{IEEEbiography}
[{\includegraphics[width=1in,height=1.25in,clip,keepaspectratio]{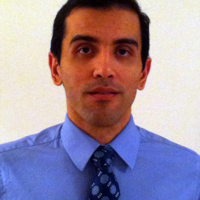}}]{Mohammad Reza Hesamzadeh}
(IEEE-SM’13) received his Docent from KTH Royal Insitute of Technology, Sweden, and his PhD from Swinburne University of Technology, Australia, in 2013 and 2010, respectively. He was a Post-Doctoral Fellow at KTH in 2010-2011 where he is currently a faculty member. He is also a Faculty Affiliate at Program on Energy and Sustainable development (PESD), Stanford University and Research Affiliate at German Institute for Economic Research (DIW Berlin). Dr Hesamzadeh is a senior memeber of IEEE, and a member of informs, IAEE and Cigre. He has won several awards for his papers in different professional events and conferences. He has published two books and numerous papers on electricity market design and analysis. He also served as editor of IEEE Transactions on Power Systems and Guest Editor of several other journals.

Dr. Hesamzadeh has been providing advice and consulting services on different energy market issues to both private and government sections over the last several years.
\end{IEEEbiography}

\begin{IEEEbiography}[{\includegraphics[width=1in,height=1.25in,clip,keepaspectratio]{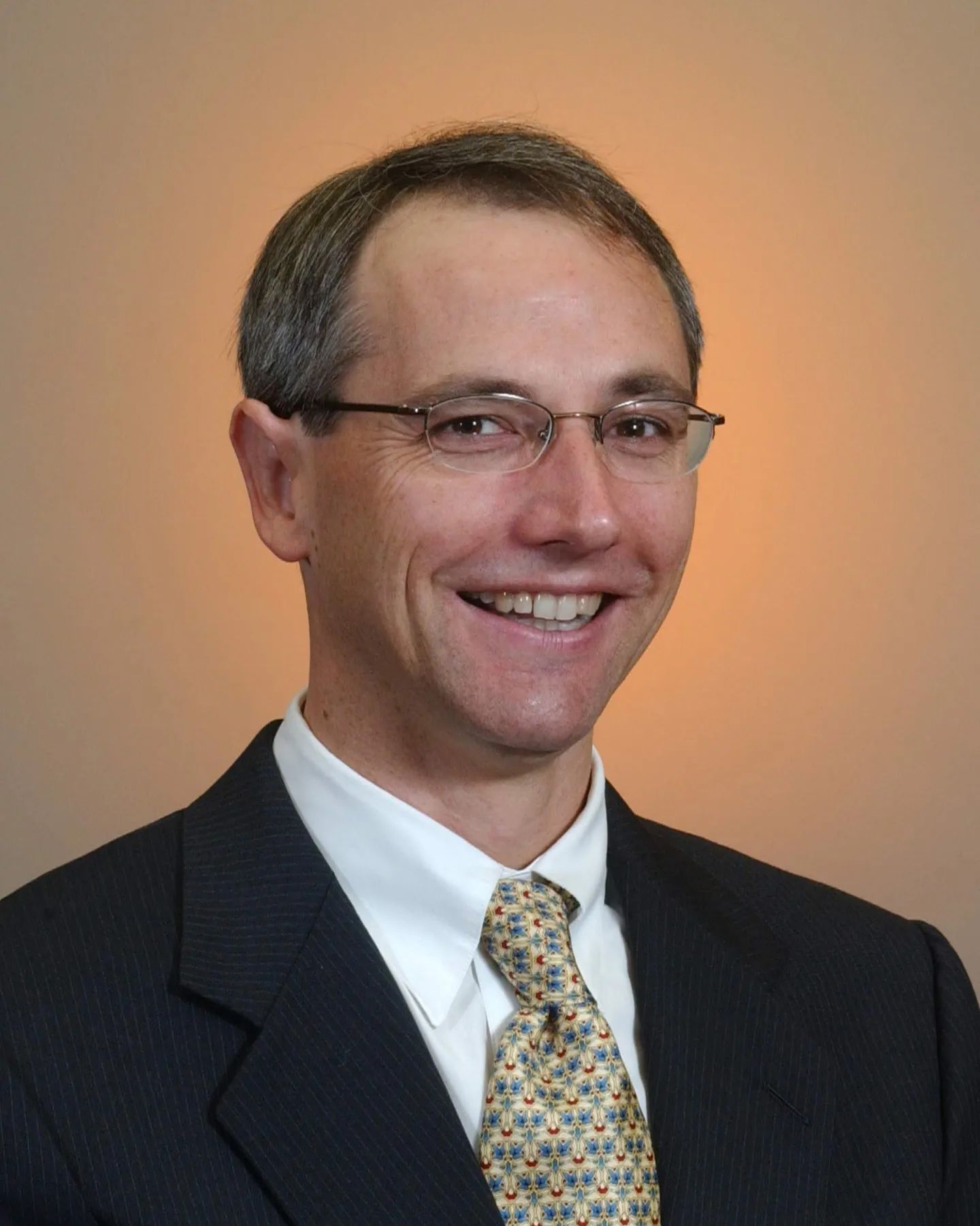}}]{Ross Baldick}
(S'90-M'91-SM'04-F'07) recieved the B.Sc. degree in Mathematics and Physics and the B.E. degree in electrical engineering from the University of Sydney, Sydney Australia and the M.S. and Ph.D. degrees in electrical engineering from the University of California, Berkeley, in 1988 and 1990 respectively.

From 1991 to 1992, he was a Post-Doctoral fellow with the Lawrence Berkeley Laboratory, Berkeley, CA. In 1992 and 1993, he was an Assistant Professor with the Worcester Polytechnic Institute, Worcester, MA. He is currently a Professor Emeritus with the Department of Electrical and Computer Engineering, The University of Texas at Austin, Austin, TX.
\end{IEEEbiography}
\end{document}